\documentclass[aps,groupedaddress]{revtex4}
\usepackage{amsmath,amssymb,graphicx,subfigure}
 \pdfoutput=1

\newcommand{\be}{\begin{eqnarray}}
\newcommand{\ee}{\end{eqnarray}}

\newcommand{\ve}{\varepsilon}
\renewcommand{\sc}{\slashchar}

\def\slashchar#1{\setbox0=\hbox{$#1$}           % set a box for #1 
   \dimen0=\wd0                                 % and get its size
   \setbox1=\hbox{/} \dimen1=\wd1               % get size of /
   \ifdim\dimen0>\dimen1                        % #1 is bigger
      \rlap{\hbox to \dimen0{\hfil/\hfil}}      % so center / in box
      #1                                        % and print #1
   \else                                        % / is bigger
      \rlap{\hbox to \dimen1{\hfil$#1$\hfil}}   % so center #1
      /                                         % and print /
   \fi}                                         %

\begin{document}

\title{Thermodynamics and quark susceptibilities:\\ a Monte-Carlo approach to the PNJL model\footnote{Work supported in part by BMBF, GSI and by the DFG cluster of excellence Origin and Structure of the Universe}}
\author{M.~Cristoforetti, T.~Hell,  B.~Klein and W.~Weise }
\affiliation{Physik-Department, Technische Universit\"at M\"unchen, D-85747 Garching, Germany}
\date {\today}

\vspace{1cm}

\begin{abstract}
The Monte-Carlo method is applied to the Polyakov-loop extended Nambu--Jona-Lasinio (PNJL) model. 
This leads beyond the saddle-point approximation in a mean-field calculation and introduces fluctuations around the mean fields.
We study the impact of fluctuations on the thermodynamics of the model, both in the case of pure gauge theory and including two quark flavors. In the two-flavor case, we calculate the second-order Taylor expansion coefficients of the thermodynamic grand canonical partition function with respect to the quark chemical potential and present a comparison with extrapolations from lattice QCD. We show that the introduction of fluctuations produces only small changes in the behavior of the order parameters for chiral symmetry restoration and the deconfinement transition. On the other hand, we find that fluctuations are necessary in order to reproduce lattice data for the flavor non-diagonal quark susceptibilities. Of particular importance are pion fields, the contribution of which is strictly zero in the saddle point approximation.
\end{abstract}  
\maketitle

\section{Introduction}
Understanding the thermodynamics of strongly interacting matter is a persistent challenge. 
The QCD phase diagram in the plane of temperature $T$ and baryonic chemical potential $\mu$ features three regions that are accessible with different strategies. Along the temperature axis with $\mu=0$, lattice QCD provides an ab-initio, non-perturbative framework, limited only by the available computing power. At asymptotically large chemical potential $\mu$, perturbative QCD is applicable. In the broad range between these extremes, models based on the symmetries and symmetry breaking patterns of QCD are useful tools for orientation. 

Different models and approaches have been developed in the last few years for this task~\cite{Fukushima:2003fm,Ratti:2005jh,Schaefer:2007pw,Braun:2009si,Braun:2009gm,Megias:2004hj}. 
A remarkably successful model in this context is the Polyakov-loop extended Nambu--Jona-Lasinio (PNJL) model. The NJL model \cite{Nambu:1961tp,Vogl:1991qt,Hatsuda:1994pi} offers a schematic, but nonetheless quite realistic, picture of the basic dynamics behind spontaneous chiral symmetry breaking: the chiral condensate as the order parameter and the pions as Goldstone bosons emerge from a chiral invariant, local four-point interaction between quarks. Quark confinement is implemented in addition by introducing the Polyakov loop as the order parameter for the confinement-deconfinement transition in the pure gauge theory and using the minimal gauge invariant coupling to quarks. This produces a dynamical entanglement of the chiral and deconfinement transitions \cite{Fukushima:2003fm,Fukushima:2003fw,Hatta:2003ga,Ratti:2005jh,Roessner:2006xn,Ratti:2007jf,Rossner:2007ik,Hell:2008cc}. Additional effects of the quark fields on the Polyakov loop potential and the phase transition at finite density have also been investigated \cite{Schaefer:2007pw,Braun:2009si,Braun:2009gm}.

A better understanding of the mechanism at the origin of these transitions requires the investigation of fluctuations in the PNJL model. In this paper we will show how 
%this improvement can be obtained 
fluctuations can be included by performing numerical simulations of the thermodynamics using standard Monte-Carlo (MC) techniques. The advantage of  this method is that it automatically incorporates fluctuations to all orders. In the present work we restrict ourselves to fluctuations of the static zero-modes which lead to an improvement beyond the saddle-point approximation.  

Since we choose a finite volume for the Monte-Carlo evaluation, in this respect our approach is similar to lattice QCD calculations. Fluctuations of particular interest are those involving Goldstone modes of both zero and finite momentum. In principle, it is necessary to take all Goldstone fluctuations into account. These fluctuations will restore the chiral symmetry in a finite volume in the absence of explicit symmetry breaking. Such effects have been investigated for example by using Renormalization Group methods \cite{Braun:2004yk,Braun:2005fj,Braun:2005gy,Braun:2008sg}. 
For sufficiently large explicit breaking of the chiral symmetry through a non-zero quark mass, as in the present work, it is legitimate to restrict the Goldstone fluctuations to the zero-modes. 
Finite-volume effects in the NJL-model have also been investigated in mean-field calculations \cite{Kiriyama:2006uh,Abreu:2006pt,Palhares:2009tf} and using sophisticated Dyson-Schwinger methods for QCD \cite{Luecker:2009bs}.

We perform our analysis in particular for the case of vanishing chemical potential where a comparison with lattice simulation results is possible. 
In this case we do not need to include diquark degrees of freedom in the model.
The behavior of the chiral condensate and the Polyakov-loop expectation value are only slightly affected by the presence of fluctuations but their effect on the susceptibilities is much more pronounced. In particular, the temperature dependence of the flavor non-diagonal second derivative of the thermodynamic grand canonical partition function with respect to quark chemical potentials is particularly sensitive to these fluctuations. For example, a quasi-particle model calculation~\cite{Bluhm:2008sc} finds a vanishing result for this susceptibility. The same result is found in a PNJL calculation using the saddle point approximation~\cite{Roessner:2009zz}. An exception to this general behavior is discussed in the NJL approach of ref.~\cite{Sasaki:2006ws} where a non-zero off-diagonal susceptibility is observed in the presence of an isovector vector coupling between quarks.
In order to account for these flavor non-diagonal susceptibilities, we include a bosonic field with the quantum numbers of the pion. In the mean field approximation, the expectation value of this field corresponds to a pion condensate. Although this pion condensate vanishes at $\mu=0$, fluctuations of this field contribute to the flavor non-diagonal Taylor expansion coefficients for the susceptibilities. 
These zero-mode fluctuations appear only in the finite-volume system and vanish in the infinite-volume limit.
In addition, fluctuations in the gauge fields also contribute to the non-diagonal susceptibilities.
These fluctuations relate to the difference of the Polyakov loop expectation value and its conjugate at finite chemical potential~\cite{Roessner:2009zz} and persist even in the infinite volume limit. Gauge field fluctuations alone are not sufficient to explain current observations on the lattice. A combination of pionic and gauge field fluctuations, on the other hand, appears to succeed. 

This paper is organized as follows: the partition function of the PNJL model is introduced in Section~\ref{sec:PNJLpf}. 
In Section~\ref{sec:mcpnjl} we discuss the model in a finite volume. An additional parameter, the ratio between the spatial and temporal extent of the Euclidean volume, is introduced, analogous to the one in finite-temperature lattice simulations. In this context we apply the Monte-Carlo method to the PNJL model.
In Section~\ref{sec:PG} results for the pure gauge case are presented, using the Monte-Carlo method with a suitable Polyakov-loop effective potential. 
In Section~\ref{sec:2nf} we study the behavior of the chiral condensate, the chiral susceptibility and the Polyakov loop for $N_f=2$ quark flavors.  Results concerning the Taylor expansion coefficients of the thermodynamic potential are shown and discussed in Section~\ref{sec:pressure}. Conclusions are presented in Section~\ref{sec:concl}.

\section{The PNJL partition function}\label{sec:PNJLpf}
The Euclidean action of the two-flavor PNJL model including finite baryon and isospin chemical potentials is given by \cite{Zhang:2006gu,Rossner:2007ik}
\be\label{eq:act}
	\mathcal{S}_{E}(\psi,\bar{\psi},\phi)&=&\int_0^{\beta}\textrm{d}\tau\int\textrm{d}^3x\Big \{\bar{\psi}(i\sc{D}+\gamma_0\tilde{\mu}-{\mathbf m})\psi+G\Big[(\bar{\psi}\psi)^2+(\bar{\psi}i\gamma_5\vec{\tau}\psi)^2\Big]\Big\}-\beta\int\textrm{d}^3x\ \mathcal{U}(\phi,\beta),
\ee
with $\beta=1/T$. Here $\psi$ is the $N_f=2$ doublet quark field, ${\mathbf m}=\textrm{diag}(m_u,m_d)$ is the quark mass matrix and the covariant derivative is
\be
	i\sc{D}=i\gamma_{\mu}(\partial^{\mu}-igA^{\mu}).
\ee
The quark chemical potential matrix $\tilde{\mu}$ is defined as
\begin{displaymath}
	\tilde{\mu}=
	\left(\begin{array}{cc}
		\mu_u&0\\
		0&\mu_d
		\end{array}
	\right).
\end{displaymath}
The Polyakov-loop effective potential $\mathcal{U}$ involves the gauge field degrees of freedom denoted by $\phi$ and models the confinement-deconfinement transition in the pure gauge theory at mean-field level. In the PNJL model quarks interact with a background color gauge field $A_4=iA_0$, where $A_0=\delta_{\mu 0}g\mathcal{A}_a^{\mu}t^a$ with the gluon fields $\mathcal{A}_a^{\mu}\in \textrm{SU}(3)_c$ and $\textrm{t}^a=\lambda^a/2$. The field $A_4$ is related to the traced Polyakov loop according to
\be\label{eq:pl}
	\Phi=\frac{1}{N_c}\textrm{tr}_c L &\textrm{with}\ L=\exp\left(i\int_0^{\beta}\textrm{d}\tau A_4\right).
\ee
In the Polyakov gauge, the matrix $L$ is given in a diagonal representation
\be 
	L=\exp(i(\phi_3\lambda_3+\phi_8\lambda_8),
	\label{eq:Ldef}
\ee
with the (diagonal) $SU(3)$ generators $\lambda_3$ and $\lambda_8$.
The dimensionless effective fields $\phi_3$ and $\phi_8$ are identified with the Euclidean gauge fields in temporal direction divided by the temperature, $A_4^{(3)}/T$ and $A_4^{(8)}/T$. These two fields parametrize the diagonal elements of $\textrm{SU}(3)_c$.

Such a description of the pure gauge thermodynamics is supposed to be valid in a limited range of $T$. It is known that, at very high temperatures, the explicit presence of transverse gluons begins to govern the dynamics. As a consequence, the validity of our approach is limited to temperatures less than about $2\ T_c$, where $T_c \sim 0.2$~GeV is the typical transition temperature scale.

In this paper we consider the ansatz for the effective potential given in \cite{Roessner:2006xn,Ratti:2006wg} motivated by the $\textrm{SU}(3)$ Haar measure which happens when integrating out six of the eight gluon fields:
\be\label{eq:effpphi}
	\frac{\mathcal{U}(\Phi,\Phi^*,T)}{T^4}=-\frac{1}{2}a(T)\Phi^*\Phi+b(T)\ln[1-6\Phi^*\Phi+4(\Phi^{*3}+\Phi^3)-3(\Phi^*\Phi)^2].
\ee
The temperature-dependent prefactors are given by
\be
	a(T)=a_0+a_1\left(\frac{T_0}{T}\right)+a_2\left(\frac{T_0}{T}\right)^2 &\textrm{and}& b(T)=b_3\left(\frac{T_0}{T}\right)^3.
\ee
As will be shown in Section~\ref{sec:PG}, the particular choice of $a(T)$ and $b(T)$ is such that we can reproduce the high-temperature behavior of thermodynamic quantities like pressure, energy and entropy density. In this way we indirectly take into account effects connected with gluonic degrees of freedom that are integrated out in the definition of our effective potential.
An additional constraint for fixing the parameters is the critical temperature of the first-order deconfinement transition in pure gauge QCD,  $T_0=270$~MeV, as given by lattice calculations, and the requirement that $\Phi^*,\Phi\rightarrow 1$ as $T\rightarrow\infty$.

Given the action (\ref{eq:act}), the partition function of our system is
\be
	\mathcal{Z}=\mathcal{N}\int\mathcal{D}\phi\mathcal{D}\psi\mathcal{D}\bar{\psi}\exp\left(-\mathcal{S}_E[\psi,\bar{\psi},\phi]\right),
\ee
where $\phi$ stands for the Polyakov loop fields $\phi_3$ and $\phi_8$.
Applying standard bosonization techniques, multiplying the partition function by the expression
\be
	\int\mathcal{D}\sigma\mathcal{D}\vec{\pi}\exp\Big[-\int_0^{1/T}\textrm{d}\tau\int\textrm{d}^3x\Big(\frac{\sigma^2+\vec{\pi}^2}{2G}\Big)\Big]
\ee
and evaluating the resulting Gaussian integral over the fermionic degrees of freedom, one finds
\be
	\mathcal{Z}=\mathcal{N}\int\mathcal{D}\phi\mathcal{D}\sigma\mathcal{D}\vec{\pi}\ \textrm{det}[S^{-1}]\exp\Big[-\frac{1}{T}\int\textrm{d}^3x\Big(\mathcal{U}(\phi,T)+\frac{\sigma^2+\vec{\pi}^2}{2G}\Big)\Big].
\ee
We write the pion field $\vec{\pi}=(\pi^1,\pi^2,\pi^3)$ in terms of $\pi^\pm=\frac{1}{\sqrt{2}}(\pi^1\pm i\pi^2)$, $\pi^0=\pi^3$ and $\tau^\pm=\frac{1}{2}(\tau^1\pm i\tau^2)$, so that
\be\nonumber
	&\vec{\tau}\cdot\vec{\pi}=\sqrt{2}(\tau^+\pi^-+\tau^-\pi^+)+\tau^3\pi^0.&
\ee
The inverse quark propagator takes the form
\begin{displaymath}
	S^{-1}=
	\left(\begin{array}{cc}
		-\sc{\partial}+(\mu_u-iA_4)\gamma_0+i\gamma_5 \pi^0-M& i\sqrt{2}\gamma_5 \pi^+\\
		i\sqrt{2}\gamma_5 \pi^-& -\sc{\partial}+(\mu_d-iA_4)\gamma_0+i\gamma_5 \pi^0-M
		\end{array}
	\right)
\end{displaymath}
with the dynamical quark mass $M=m_0-\sigma$ generated by the scalar field $\sigma<0$. We work in the isospin symmetric limit with $m_0=m_u=m_d$ for convenience. This scalar field is related to the chiral (quark) condensate by $\sigma=G\langle\bar{\psi}\psi\rangle$.
Equivalently, the partition function can be rewritten as
\be\label{eq:pfunc}
	\mathcal{Z}&=&\mathcal{N}\int\mathcal{D}\phi\mathcal{D}\sigma\mathcal{D}\vec{\pi}\exp\left(-\mathcal{S}[\sigma,\vec{\pi},A_4]\right)\nonumber\\
	&=&\mathcal{N}\int\mathcal{D}\phi\mathcal{D}\sigma\mathcal{D}\vec{\pi}\exp\left[\frac{1}{2}\textrm{Tr}\ln[S^{-1}]-\frac{1}{T}\int\textrm{d}^3x\Big(\mathcal{U}(\phi,T)+\frac{\sigma^2+\vec{\pi}^2}{2G}\Big)\right].
\ee
The standard path for the evaluation of this partition function is to first consider the mean-field limit, where only the classical trajectory in the path integral is taken into account. In this approximation, the partition function becomes an ordinary multi-dimensional integral. This mean-field partition function is evaluated using the saddle-point approximation, taking into account only the maximum contribution to the partition function. To evaluate this contribution, the action is minimized by variation of the classical fields. This approximation becomes exact only in the infinite-volume limit, which is not reached in lattice simulations. In the present paper we will consider a different approach.

\section{PNJL model in a finite volume}\label{sec:mcpnjl}
\subsection{Finite-Volume partition function}
In the present calculation we perform a step beyond mean-field approximation by including fluctuations of the zero modes of the relevant fields. This is admittedly only part of all possible field fluctuations, but it represents nevertheless an improvement with respect to the standard mean-field calculation.
These zero-mode fluctuations can be introduced considering a system defined in a finite volume $V$.

The partition function in momentum space is written as
\be\label{eq:pfuncvi}
	\mathcal{Z}&=&\int\mathcal{D}\phi\mathcal{D}\sigma\mathcal{D}\vec{\pi}\exp\Big[\frac{V}{T}\Big(\frac{1}{2}\sum_n\sum_{\vec{p}}\textrm{Tr}\ln[S^{-1}(i\omega_n,\vec{p})]-\mathcal{U}(\phi,T)-\frac{\sigma^2+\vec{\pi}^2}{2G}\Big)\Big]
\ee
where $\omega_n=(2n+1)\pi T$ are the Matsubara frequencies. 
Given the inverse quark propagator defined by
\begin{displaymath}
	S^{-1}(i\omega_n,\vec{p}\,)=
	\left(\begin{array}{cc}
		(i\omega_n+\mu_u-iA_4)\gamma_0+i\gamma_5\pi^0-\vec{\gamma}\cdot\vec{p}-M& i\sqrt{2}\gamma_5 \pi^+\\
		i\sqrt{2}\gamma_5 \pi^-& (i\omega_n+\mu_d-iA_4)\gamma_0+i\gamma_5\pi^0-\vec{\gamma}\cdot\vec{p}-M
		\end{array}
	\right)
\end{displaymath}
the fermionic determinant can be evaluated by diagonalization and one finds
\be
	\frac{1}{2}\textrm{Tr}\ln[S^{-1}]=-2 N_f\int \frac{\textrm{d}^3p}{(2\pi)^3}\sum_j\left\{T\ln\left[1+e^{-E_j/T}\right]+\frac{1}{2}\Delta E_j\right\}
\ee
with twelve quasiparticle energies
\be
	E_{1,2}=\sqrt{(\ve(p)+\mu_I)^2+2\pi^+\pi^-}\pm(\mu-2iA_8/\sqrt{3}), && E_{3,4}=\sqrt{(\ve(p)-\mu_I)^2+2\pi^+\pi^-}\pm(\mu-2iA_8/\sqrt{3}), \nonumber\\
	E_{5,6}=\sqrt{(\ve(p)+\mu_I)^2+2\pi^+\pi^-}\pm(\mu+i(A_3+A_8/\sqrt{3})), && E_{7,8}=\sqrt{(\ve(p)-\mu_I)^2+2\pi^+\pi^-}\pm(\mu+i(A_3+A_8/\sqrt{3})), \nonumber\\
	E_{9,10}=\sqrt{(\ve(p)+\mu_I)^2+2\pi^+\pi^-}\pm(\mu-i(A_3-A_8/\sqrt{3})), && E_{11,12}=\sqrt{(\ve(p)-\mu_I)^2+2\pi^+\pi^-}\pm(\mu-i(A_3-A_8/\sqrt{3})), \nonumber
\ee
where we have introduced isoscalar and isovector chemical potentials, $\mu=(\mu_u+\mu_d)/2$, $\mu_I=(\mu_u-\mu_d)/2$ and $\ve(p)=\sqrt{p^2+\pi_0^2+(m_0-\sigma)^2}$.  The energy difference is defined as the difference between the quasiparticle energy and the energy of a free fermion, $\ve_0=\sqrt{p^2+m_0^2}\pm\mu_I$: $\Delta E_j=E_j-\ve_0\mp\mu$.

For the volumes considered in the present work, effects from discretization of momenta turn out to be negligible. In order to simplify the numerical evaluation of the partition function, we have therefore substituted an integral for the sum over the discrete three-momenta. This approximation becomes invalid in the limit of small volume size and small quark masses and it is not applicable in general. 

The presence of a volume factor $V$ in the exponent of Eq.~(\ref{eq:pfuncvi}) makes it possible to compute the full partition function in mean-field approximation using Monte-Carlo techniques, as explained in the following section. In this way we consider not only the saddle-point contributions, but also configurations that correspond to fluctuations around the minima of the action. 
 %Obviously, at this level the fluctuations are of purely thermal origin, and in the infinite volume limit they disappear. 
The zero-mode fluctuations of the pion fields which appear at this level will disappear in the infinite-volume limit.   
%Nevertheless we will see in the last section of the paper that a different interpretation of this volume can be given and from this new point of view fluctuations survive in the infinite volume limit. Actually, this different interpretation will be essential in the understanding of our result concerning the second non-diagonal Taylor expansion coefficient of the thermodynamic potential.
Additionally, contributions from the gauge fields which are related to the difference between the Polyakov loop and its conjugate also contribute to flavor non-diagonal susceptibilities. These contributions survive in the infinite-volume limit.

The size of the volume is now specified according to the conventions adopted in lattice calculations. For a fixed extension of the lattice in the Euclidean time direction, the temperature is set by the lattice spacing $a$, and the volume size is related to the temperature:
\be
	a=\frac{1}{N_t T}& \rightarrow &V=N_s^3 a^3=\frac{N_s^3}{N_t^3 T^3},
\ee
where $N_t$ is the number of lattice sites in the Euclidean time direction, and $N_s$ is the number of lattice sites in the space direction.
It follows that 
\be\label{eq:vk}
	V = k/T^3,
\ee
where different values of $k=(N_s/N_t)^3$ will be chosen for our purpose. On the lattice the parameter $k$ must be sufficiently large in order to reduce finite volume effects while keeping the system at a specified temperature. On the other hand, for increasing values of $k$ the necessary computing time grows exponentially at fixed temperature and simulations for very large volume become less feasible. At present, $N_s/N_t=4$  ($k=64$) is the largest value used in lattice simulations \cite{Gavai:2008zr,Allton:2005gk}.
In the MC-simulation of the PNJL model, the aim is to understand the volume dependence of thermodynamic quantities. For this reason we perform calculations at different values of $k$, including $k=64$ which corresponds to the typical lattice simulation volume.

\subsection{Monte-Carlo method}\label{sec:MC}
In Euclidean quantum field theory, the expectation value of an observable $\mathcal{O}$ is given by
\be\label{eq:expv}
	\langle\mathcal{O}\rangle=\frac{\int\mathcal{D}\varphi\,\mathcal{O}(\varphi)e^{-\mathcal{S}[\varphi]}}{\int\mathcal{D}\varphi\,e^{-\mathcal{S}[\varphi]}}.
\ee
where $\varphi$ stands for the set of relevant field variables. A statistical method is used in order to select those configurations of the fields $\varphi$ which give a significant contribution to (\ref{eq:expv}).
An efficient way of computing the ensemble average consists of generating a sequence of configurations with a probability distribution given by $\exp(-\mathcal{S}[\varphi])$. This technique is called ``importance sampling". If the sequence generated constitutes a representative set of configurations, then the ensemble average $\langle\mathcal{O}\rangle$ will be given approximately by the sum
\be
	\langle\mathcal{O}\rangle\approx\frac{1}{N}\sum_{i=1}^N\mathcal{O}({\varphi}_i),
\ee
where ${\varphi}_i\ (i=1,...,N)$ denote the configurations generated.

There are different techniques to generate the relevant ensemble of configurations. In this paper we adopt the Metropolis method.
The algorithm is simple: starting from a random configuration $C$, a new configuration $C'$ is generated with acceptance
\be
	p=\textrm{min}\{1,\exp(-\mathcal{S}[C'])/\exp(-\mathcal{S}[C])\}.
\ee
The new configuration will be accepted if the action has decreased; if it has increased, it will be accepted with probability $\exp(-\mathcal{S}[C'])/\exp(-\mathcal{S}[C])$. 
If we use a cooling procedure and accept only those configurations that decrease the action, the chain of configurations generated in this way ends up with a configuration that corresponds to the saddle point result, i.e. with a configuration that minimizes the action. 
%In particular, if we accept only configurations which lower the action (cooling procedure) we will find, at the end of our configuration chain, the mean-field result i.e. the configuration that minimizes the total action. 
%
In the next sections we will see how this Monte-Carlo evaluation of the PNJL model works both in the pure gauge sector -- i.e. with the Polyakov loop potential -- and in the case with $N_f=2$ quark flavors.

\section{Monte-Carlo evaluation of the Polyakov Loop sector}\label{sec:PG}

The effective potential for the Polyakov loop in Eq.~(\ref{eq:effpphi}) is written as a function of the traced Polyakov loop $\Phi$. In this section we study the properties of this potential in terms of the fundamental gauge fields $\phi_3=\frac{A_4^{(3)}}{T}$ and $\phi_8=\frac{A^{(8)}_4}{T}$.  Using the definition of the Polyakov loop given in Eq.~(\ref{eq:pl}), and the representation (\ref{eq:Ldef})
\be
	\Phi(\phi_3,\phi_8)=\frac{1}{3}\Big(e^{i(\phi_3+\frac{\phi_8}{\sqrt{3}})}+e^{i(-\phi_3+\frac{\phi_8}{\sqrt{3}T})}+e^{i\frac{2\phi_8}{\sqrt{3}}}\Big),
\ee
we obtain for the effective potential (\ref{eq:effpphi}) expressed in terms of $\phi_3$ and $\phi_8$:
\be\label{eq:effpA}
	\frac{\mathcal{U}(\phi_3,\phi_8,T)}{T^4}&=&-\frac{1}{18}a(T)\left[3+2\cos(2 \phi_3)+4\cos(\phi_3)\cos(\sqrt{3}\phi_8)\right]\nonumber\\
	&&+b(T)\ln\left[\frac{16}{27}\left(\cos(\phi_3)-\cos(\sqrt{3}\phi_8)\right)^2\sin(\phi_3)^2\right].
\ee
\begin{figure}[ht!]
       \centering
        \subfigure{\includegraphics[width=.45\textwidth]{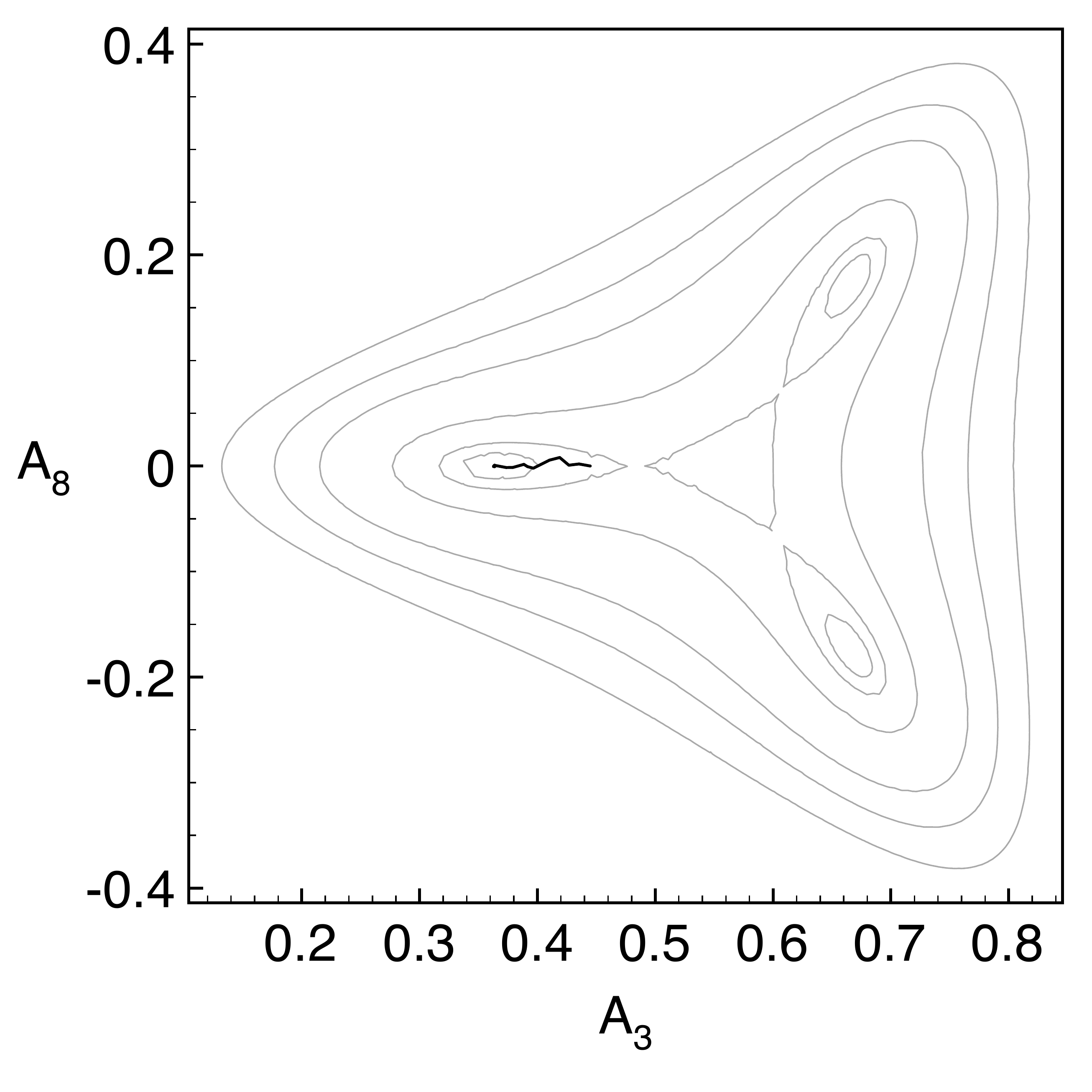}}\hspace{5mm}%
       \subfigure{\includegraphics[width=.45\textwidth]{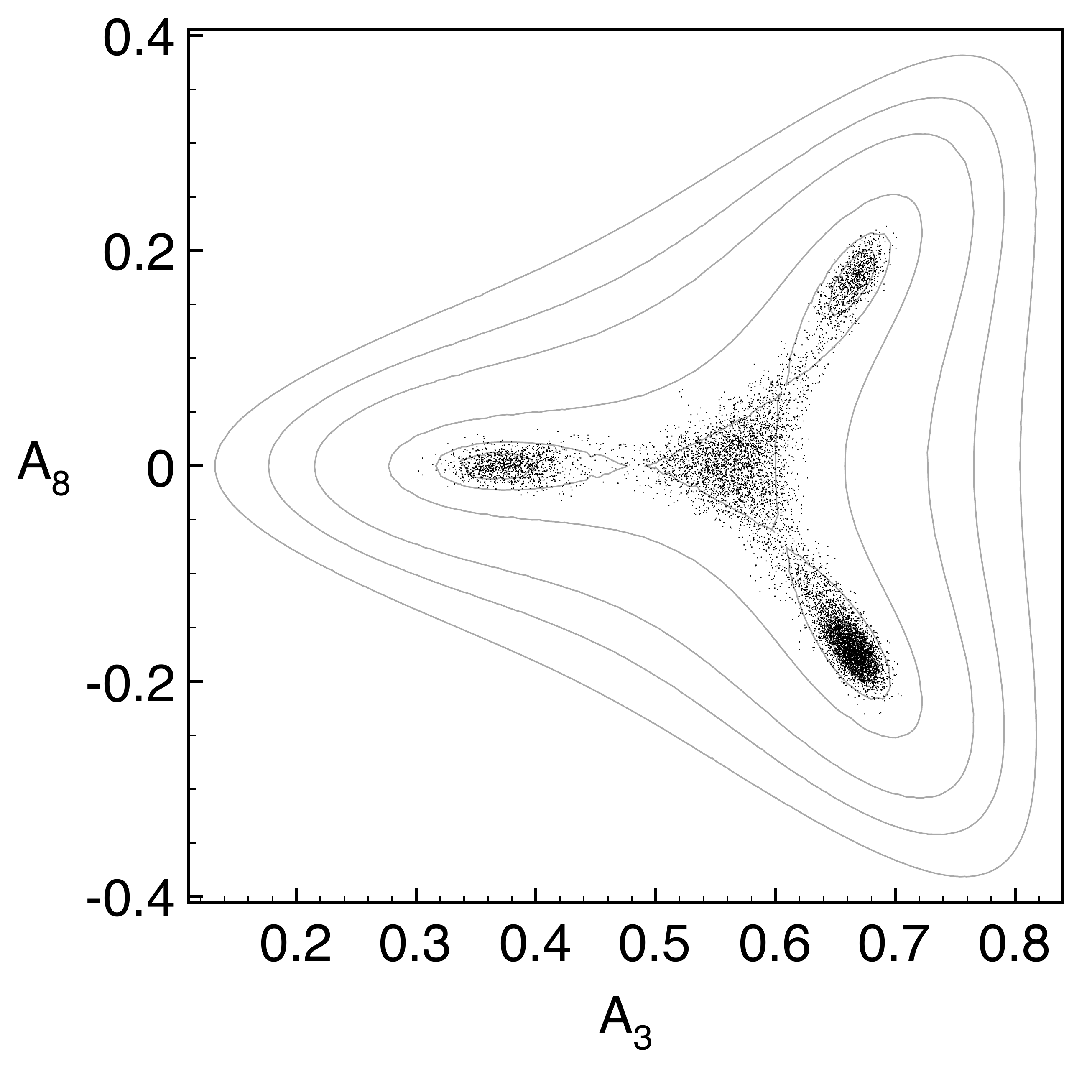}}
      \caption{Comparison between the trajectory generated by a cooling algorithm (left) and the field configurations produced by the Metropolis algorithm (right) applied to the Polyakov loop effective potential at temperature $T=T_c=0.27$ GeV. The lines represent equipotential curves of ${\mathcal U}$ (Eq.~(\ref{eq:effpphi})) in the $A_3$-$A_8$-plane.}\label{fig:CoolvsMetr}
\end{figure}

In this section the Monte-Carlo approach is applied to the sampling of the Polyakov loop. Although contributions from fluctuations are small in this case, it is instructive to illustrate  how the approach works.
We start from a generic configuration fixing the $A^{(3)}_4$ and $A^{(8)}_4$ fields at a given temperature. Using the Monte-Carlo method, we generate points in configuration space close to the minimum of the action. 
In particular, applying cooling, we reach exactly the minimum. Fig.~\ref{fig:CoolvsMetr} illustrates the difference between the trajectories generated by the cooling procedure and the field samples produced by the Metropolis algorithm at the critical temperature $T=T_0=0.27$ GeV. For cooling (left figure) the action must decrease, and in a few steps one of the three minima of the potential is reached. On the other hand, using the full Metropolis algorithm, the action can also increase occasionally, and as a consequence the space of the available configurations is much larger (right figure). 

\begin{figure}[ht!]
        \centering
        \subfigure{\includegraphics[width=.46\textwidth]{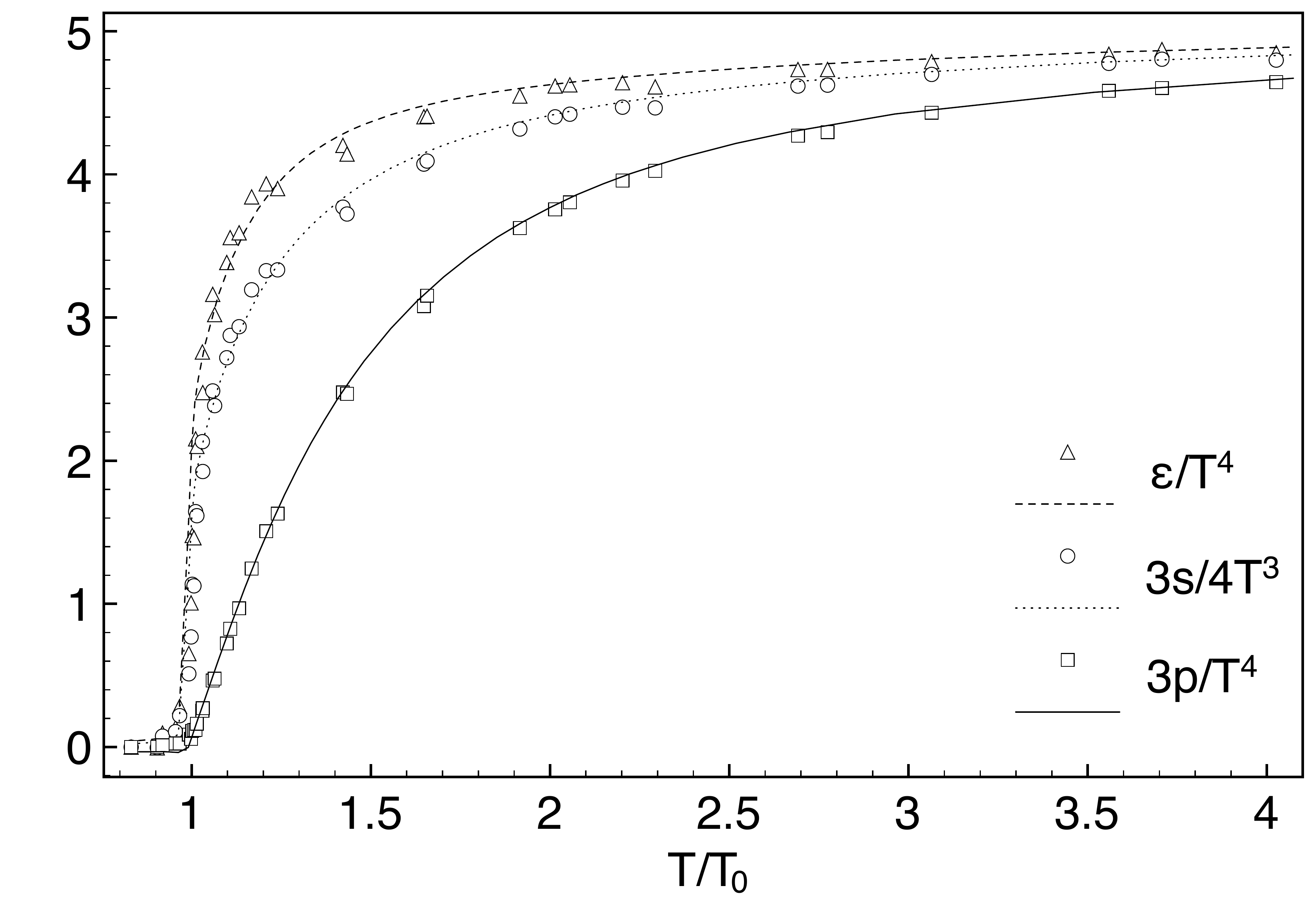}}\hspace{5mm}%
      \subfigure{\includegraphics[width=.46\textwidth]{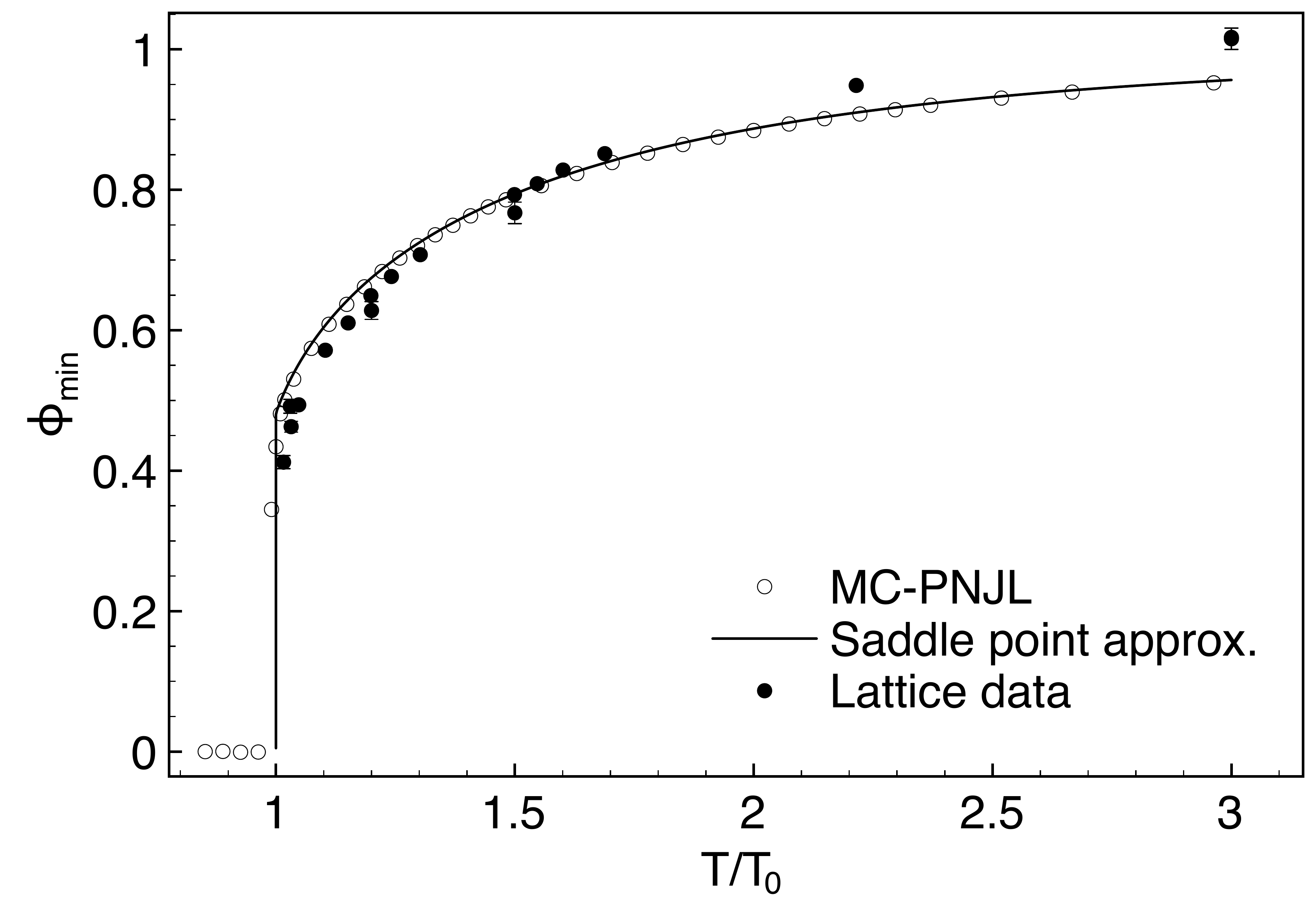}}
      \caption{Left: Results for pressure $p$, energy density $\varepsilon$ and entropy density $s$ from the Monte-Carlo evaluation of the Polyakov loop effective potential fitted to lattice results (lines), compared to lattice QCD data from \cite{Boyd:1996bx} (open symbols). Right: Comparison of the Polyakov loop expectation value from the saddle point approximation (solid line) and the Monte-Carlo evaluation (open points) with lattice QCD data taken from \cite{Kaczmarek:2002mc} (solid points).}\label{fig:peePG}
\end{figure}

To fix the parameters of the effective potential, we have computed the pressure, energy and entropy density for the pure gauge theory at different temperatures, including fluctuations, and chosen parameter values to fit the available lattice data (Fig.~\ref{fig:peePG} left). It is found that only the value of the parameter $b_3$ differs slightly from the one obtained by mean-field calculations in the saddle point approximation. 
With the smallest value for the parameter $k$ used in the calculation, which leads to the largest fluctuations, we find $b_3=-1.64$, compared to $b_3=-1.68$ from the saddle point approximation. 
% moving from $b_3=-1.68$ to $b_3=-1.64$, using the smallest value of our parameter $k$ (which leads to larger fluctuations). 
In the right panel of Fig.~\ref{fig:peePG} we compare the expectation value for the Polyakov loop obtained 
%using the mean-field method 
in the saddle point approximation (line) with our results (open points) and the lattice data (points). There is practically no difference between the two methods, cooling versus Metropolis: both agree well with lattice data.  
Once we have established how the Monte-Carlo approach works in the case of Polyakov loop dynamics, we can move on to the more interesting case with two quark flavors.

\section{Monte-Carlo approach to the two-flavor PNJL model}\label{sec:2nf}
The starting point for studying the thermodynamics for $N_f=2$ quark flavors is the partition function (\ref{eq:pfuncvi}). The degrees of freedom in this case are the $A^{(3)}_4$ and $A^{(8)}_4$ components of the gauge field, and the bosonic field variables $\sigma$ and $\vec{\pi}$. In the evaluation of the path integral, we need to fix the volume $V$ as a function of the temperature. 
%From (\ref{eq:vk}) it means that we only have to fix the value of $k$. 
Looking back at Eq.~(\ref{eq:vk}), we see that this requires fixing the dimensionless index $LT=k^{1/3}$ of the Euclidean volume. 
In this work we consider six different choices of $k$. The value $k=64$ corresponds to the largest one currently used in lattice simulations. In addition we consider $k=125,250,500,1000,2500$. The ratio between the smallest and the largest $k$ is~$\sim40$. In this way we can study systematically the dependence of the observables on the volume size at fixed temperature $T$. In the NJL sector of the model we also need to specify the current quark mass $m_0$, the coupling constant $G$ and the three-momentum cut-off $\Lambda$. The parameters used here are the ones of Refs.~\cite{Rossner:2007ik,Roessner:2006xn}:
\be
	m_0=5.5\ \textrm{MeV}, & G=10.1\ \textrm{GeV}^{-1}, & \Lambda=650\ \textrm{MeV}.\nonumber
\ee

\subsection{Chiral and deconfinement transitions}
\begin{figure}[ht!]
        \centering
        \includegraphics[width=.6\textwidth]{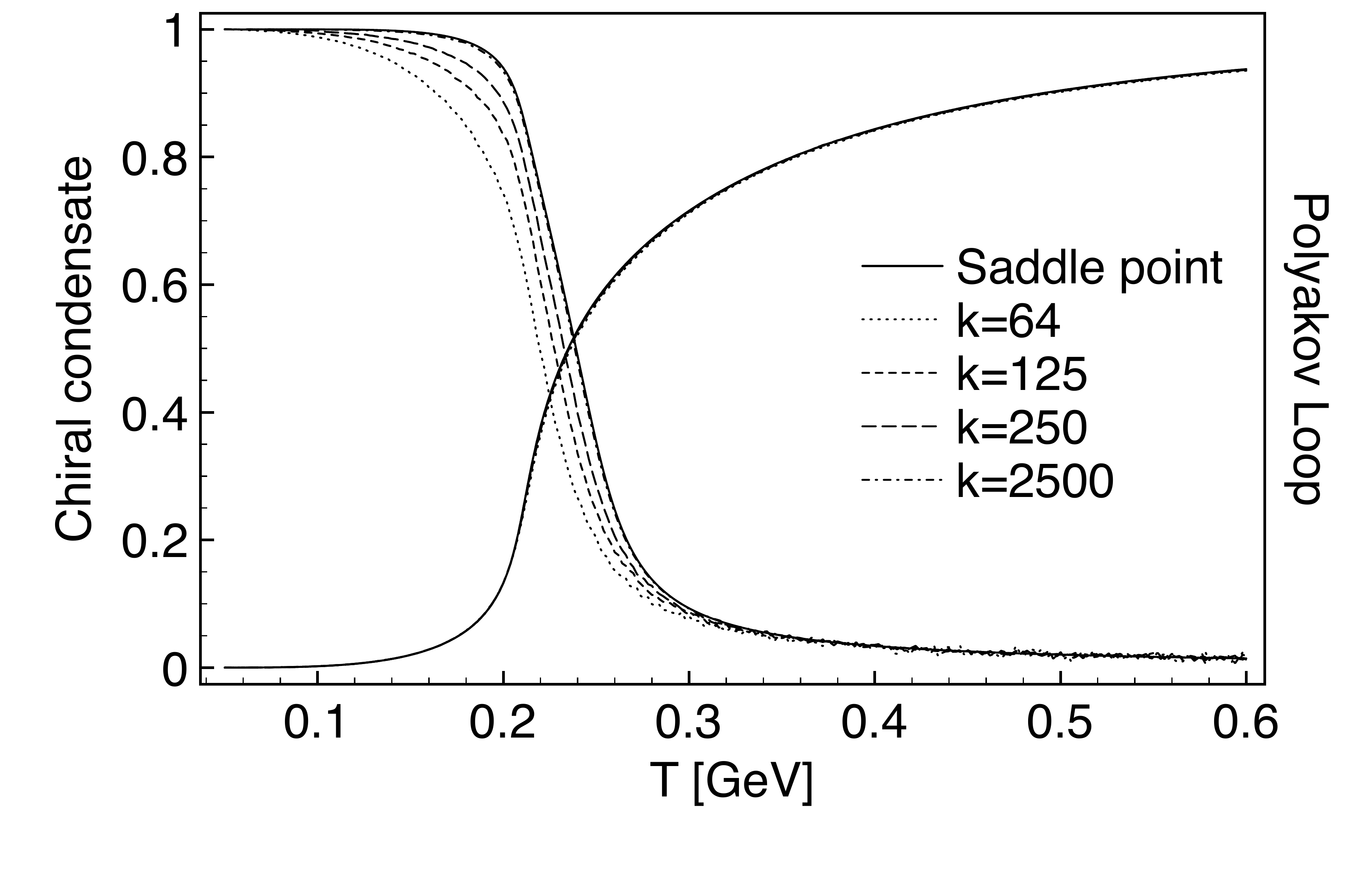}
      \caption{Dependence of the chiral condensate, $\langle\bar{\psi}\psi\rangle_T/\langle\bar{\psi}\psi\rangle_0$, and of the Polyakov loop expectation value $\langle\Phi\rangle$ on the parameter $k=(LT)^3$ in a finite volume. Deviations from the mean-field result in the infinite-volume limit are manifest only for the chiral condensate. The behavior of the Polyakov loop is completely unchanged.}\label{fig:comop}
\end{figure}

The principal aim of this work is to contribute to the investigation of the QCD phase diagram. In this section we study how the chiral and deconfinement transitions are affected by the introduction of fluctuations around the mean field in our Monte-Carlo PNJL approach. This is achieved by evaluating the chiral condensate and the Polyakov loop expectation value for different volumes and comparing with the mean-field result in saddle point approximation. This comparison is presented in Fig.~\ref{fig:comop}. The presence of fluctuations does obviously not modify the behavior of the Polyakov loop expectation value; the four different sets of data overlap perfectly. For the chiral condensate below the critical temperature, we notice that there is in fact a non-trivial dependence on the temperature: the expectation value of the  field $\sigma$ starts to decreases earlier for smaller $k$. This reflects a volume dependence that moves the chiral transition temperature from $T_\sigma=254$ MeV for $k=2500$ to $T_\sigma=222$ MeV for $k=64$, as deduced from the following analysis of the chiral susceptibility.

\subsection{Chiral susceptibility}
The chiral susceptibility is sensitive to fluctuations in the fields and therefore provides a good test for the Monte-Carlo evaluation of the PNJL model. In the infinite-volume limit, $V\rightarrow\infty$, the Monte-Carlo calculation should recover the saddle-point result. To perform this comparison, we also calculate the chiral susceptibility in the saddle-point approximation.

\begin{figure}[ht!]
        \centering
        \includegraphics[width=.6\textwidth]{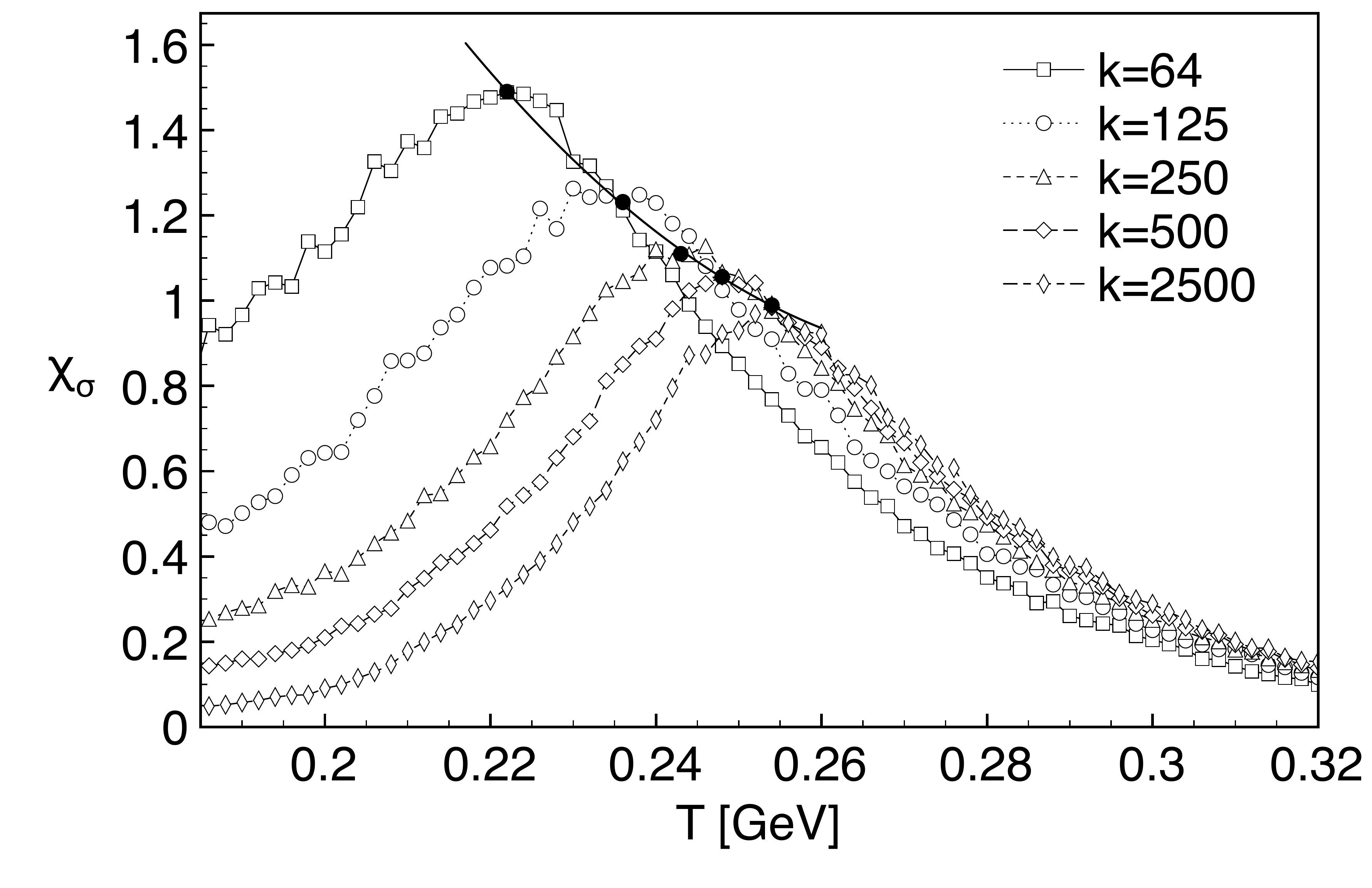}
      \caption{The chiral susceptibility as a function of temperature for different volume aspect ratios. We use the peak position as a measure for the chiral transition temperature $T_\sigma$. The solid line shows the trajectory of the chiral transition temperature as it rises with increasing volume.}\label{fig:chsu}
\end{figure}

The chiral susceptibility is defined as
\be
	\chi_{\sigma}=\frac{T}{V}\frac{\partial^2}{\partial m_0^2}\ln \mathcal{Z}(m_0, T),
\ee
in terms of the partition function $\mathcal{Z}$ of Eqns.~(\ref{eq:pfunc}),(\ref{eq:pfuncvi}). The second derivative is taken with respect to the quark mass $m_0$. Performing these derivatives leads to 
\be\label{eq:chsu}
	\chi_{\sigma}&=&\frac{V}{T}\Big[\frac{1}{\mathcal{Z}(m_0,T)}\int\mathcal{D}\sigma\mathcal{D}\vec{\pi}\mathcal{D}A \left(\frac{\partial\ln\det S^{-1}(m_0,T,\sigma,\vec{\pi},A)}{\partial m_0}\right)^2 e^{\mathcal{-S}[\sigma,\vec{\pi},A]}\nonumber\\
	&&-\Big(\frac{1}{\mathcal{Z}(m_0,T)}\int\mathcal{D}\sigma\mathcal{D}\vec{\pi}\mathcal{D}A \frac{\partial\ln\det S^{-1}(m_0,T,\sigma,\vec{\pi},A)}{\partial m_0} e^{\mathcal{-S}[\sigma,\vec{\pi},A]}\Big)^2\Big]\nonumber\\
	&&+\frac{1}{\mathcal{Z}(m_0,T)}\int\mathcal{D}\sigma\mathcal{D}\vec{\pi}\mathcal{D}A \frac{\partial^2\ln\det S^{-1}(m_0,T,\sigma,\vec{\pi},A)}{\partial m_0^2} e^{\mathcal{-S}[\sigma,\vec{\pi},A]}\nonumber\\
	&=&\frac{V}{T}\Big[\Big\langle\left(\frac{\partial\ln\det S^{-1}(m_0,T,\sigma,\vec{\pi},A)}{\partial m_0}\right)^2\Big\rangle-\Big\langle\frac{\partial\ln\det S^{-1}(m_0,T,\sigma,\vec{\pi},A)}{\partial m_0}\Big\rangle^2\Big]\nonumber\\
	&&+\Big\langle\frac{\partial^2\ln\det S^{-1}(m_0,T,\sigma,\vec{\pi},A)}{\partial m_0^2}\Big\rangle.
\ee
This expression can now be evaluated using the Monte-Carlo algorithm. We use the position of the peak in the chiral susceptibility as a measure for the chiral transition temperature $T_\sigma$. The results presented in Fig.~\ref{fig:chsu} show that $T_\sigma$ moves toward its saddle point limit, $T_\sigma\approx 254$ MeV (black points). 

\begin{figure}[ht!]
        \centering
        \includegraphics[width=.6\textwidth]{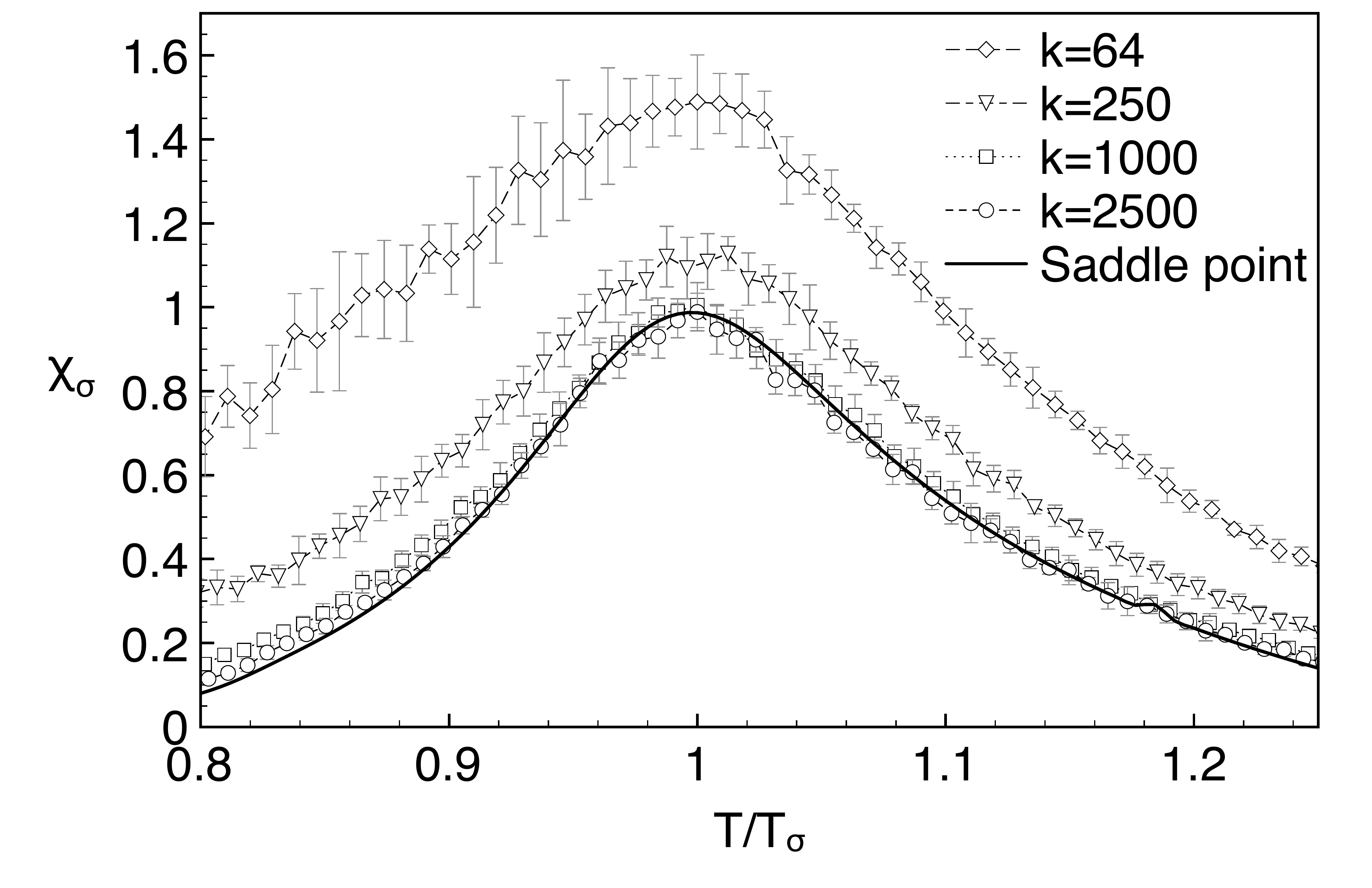}
      \caption{Chiral susceptibility as a function of $T/T_\sigma$: With increasing volume the Monte-Carlo results (open symbols) approach the saddle-point mean field result (solid curve).}\label{fig:chsuSP}
\end{figure}

The mean-field numerical result is obtained by evaluating the thermodynamic potential in saddle-point approximation for current quark masses $m_0$ in the range from $1$ to $10$~MeV and for different temperatures. The thermodynamic potential is then interpolated and the second derivative is calculated numerically. The result is represented by the solid curve in Fig.~\ref{fig:chsuSP}.
The Monte-Carlo calculation approaches the mean-field limit when the volume is increased at fixed temperature. 

The role of fluctuations in the Monte-Carlo calculation of the chiral susceptibility can be understood from Eq.~(\ref{eq:chsu}).
%actually we have for the term in squared bracket, the disconnected contribution, we have 
The disconnected contribution (the term in the square brackets), vanishes in the infinite-volume limit since fluctuations of the mean field contribute as $\sim1/V$:
\be
	\langle \bullet^2\rangle-\langle \bullet\rangle^2\rightarrow0 &\textrm{as}&  V\rightarrow\infty,
\ee
Because the prefactor $V/T$ in Eq.~(\ref{eq:chsu}) compensates the volume dependence of the leading fluctuation contributions, a finite susceptibility results also in the limit $V\rightarrow\infty$. Since additional contributions of fluctuations in the mean fields are of higher order in $1/V$, the saddle point approximation becomes exact in this limit.

\section{Non-zero quark chemical potentials: taylor expansion}\label{sec:pressure}
Dealing with non-zero quark chemical potentials $\mu_q$ in lattice QCD thermodynamics is notoriously difficult because of the well-known fermion sign problem. A possible way of overcoming this problem is the Taylor-expansion approach.
%The Taylor expansion approach to QCD thermodynamics at non-zero quark chemical potential represents one of the possibilities to overcome the sign problem on the lattice. 
Instead of performing an explicit calculation at $\mu_q\neq0$, the thermodynamic potential is expanded in a Taylor series in $\mu_q/T$ around zero chemical potential,
\be
	\Omega(T,\mu)=\frac{1}{VT^3}\ln\mathcal{Z}=\sum_{i,j=0}^{\infty}\chi_{ij}(T)\left(\frac{\mu_u}{T}\right)^i\left(\frac{\mu_d}{T}\right)^j,
\ee
with
\be
	\chi_{ij}(T)=\frac{1}{i!j!}\left.\frac{\partial^{i+j}\Omega}{\partial(\mu_u/T)^i\partial(\mu_d/T)^j}\right|_{\mu_u=\mu_d=0},
\ee
where only even terms survive due to $CP$ symmetry. The coefficients $\chi_{ij}(T)$ are evaluated at $\mu_q=0$.

The comparison between lattice data and Monte-Carlo calculations for these coefficients in the PNJL model represent an important test of this model. In particular the flavor non-diagonal coefficient $\chi_{11}$ that vanishes in the saddle point approximation is of interest in this context: it is necessary to take fluctuations of the mean field into account in order to obtain a non-vanishing result for $\chi_{11}$. Since the strength of fluctuations depends on the volume, we again evaluate the Taylor coefficients for different volume sizes at each temperature, i.e. for different values of the parameter $k$.  
%In particular, the second non-diagonal coefficient is the most important, because, as explained in the introduction, to obtain a non-vanishing result for $c_2^{ud}$ we need to take into account fluctuations of the fields. We know that the size of fluctuations in our model depend on the size of our correlation-volume, therefore, as before, we evaluate the Taylor coefficients for different values of the parameter $k$.

\subsection{Second order Taylor expansion coefficients and susceptibilities}\label{sec:sotc}
\begin{figure}[ht!]
       \centering
    \includegraphics[width=.6\textwidth]{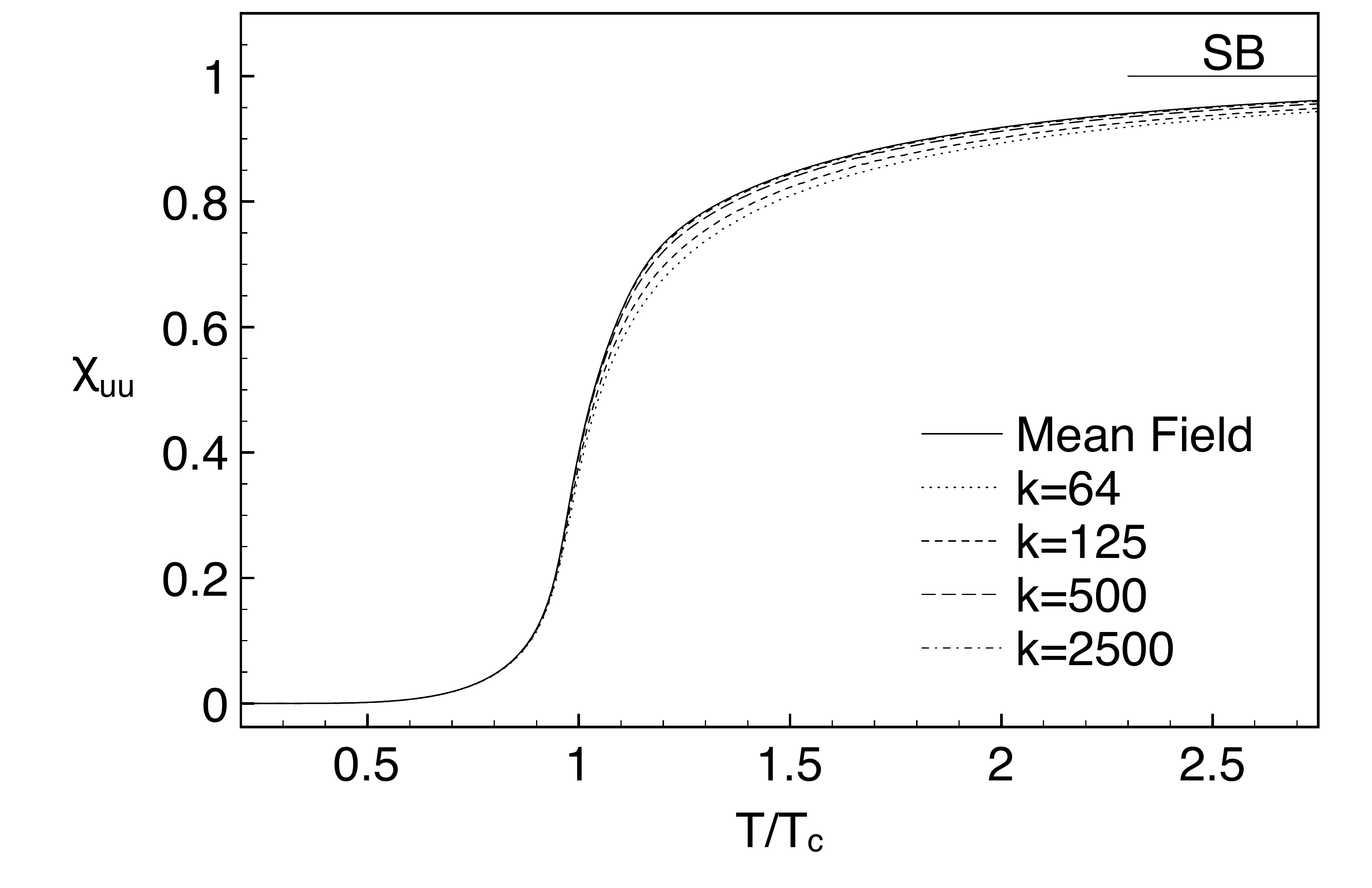}
      \caption{Flavor-diagonal second order expansion coefficient $\chi_{uu}=2\chi_{20}$ for different ratios $k$  compared with the saddle point approximation. The Stefan-Boltzmann limit is marked by SB.}\label{fig:c2uu}
\end{figure}

The first derivative of the logarithm of the partition function gives
\be\label{eq:c2uda}
	\frac{\partial\ln\mathcal{Z}(T,\mu_u,\mu_d)}{\partial\mu_q}&=&\frac{\partial}{\partial\mu_q}\ln\int\mathcal{D}\sigma\mathcal{D}\vec{\pi}\mathcal{D}A\exp\Big[\frac{V}{T}\ln\det S^{-1}(T,\mu_u,\mu_d,\sigma,\vec{\pi},A)-\mathcal{S}_g[A]\Big]\nonumber\\
	&=&\frac{1}{\mathcal{Z}(T,\mu_u,\mu_d)}\frac{V}{T}\int\mathcal{D}\sigma\mathcal{D}\vec{\pi}\mathcal{D}A\frac{\partial\ln\det S^{-1}(T,\mu_u,\mu_d,\sigma,\vec{\pi},A)}{\partial\mu_q}e^{-\mathcal{S}[T,\mu_u,\mu_d,\sigma,\vec{\pi},A]}\nonumber\\
	&=&\frac{V}{T}\Big\langle\frac{\partial\ln\det S^{-1}(T,\mu_u,\mu_d,\sigma,\vec{\pi},A)}{\partial\mu_q}\Big\rangle.
\ee
Proceeding in the same way for the second derivative, we obtain the coefficients $\chi_{uu}$ and $\chi_{ud}$ (quark susceptibilities)
\be\label{eq:c2ud}
	\chi_{uq}&=& \frac{1}{VT}\frac{\partial^2}{\partial\mu_u\partial\mu_q}\ln\mathcal{Z}= \frac{T^2}{VT^3}\Bigg(\frac{V}{T}\left\langle\frac{\partial^2}{\partial\mu_u\partial\mu_q}\ln\det S^{-1}(T,\mu_u,\mu_d,\sigma,\vec{\pi},A)\right\rangle \nonumber\\
	&&+\left(\frac{V}{T}\right)^2\left\langle\left(\frac{\partial}{\partial\mu_u}\ln\det S^{-1}(T,\mu_u,\mu_d,\sigma,\vec{\pi},A)\right)^2\right\rangle-\left(\frac{V}{T}\right)^2\left\langle\frac{\partial}{\partial\mu_u}\ln\det S^{-1}(T,\mu_u,\mu_d,\sigma,\vec{\pi},A)\right\rangle^2 \Bigg).
\ee
Consider first the flavor-diagonal susceptibility $\chi_{20}=\frac{1}{2}\chi_{uu}$. A comparison between Monte-Carlo results and the saddle point approximation for $\chi_{uu}$ is shown in Fig.~\ref{fig:c2uu} for different volumes as a function of the temperature. In this case the contributions of fluctuations are evidently small, reflecting the fact that $\chi_{uu}$ is governed by the non-vanishing quark condensate and the $A_3$ component of the gauge field.

The behaviour of the flavor non-diagonal coefficient $\chi_{11}=\chi_{ud}$, on the other hand, is quite different. It vanishes in the saddle point approximation whereas lattice QCD clearly displays a non-zero signal for this quantity around $T_c$.

For the detailed evaluation of $\chi_{ud}$ we return to the expressions, Eqs.~(\ref{eq:c2uda}),(\ref{eq:c2ud}), and evaluate derivatives of the fermion determinant:
\be\label{eq:c2udfd}
	&&\frac{\partial}{\partial\mu_u}\ln\det S^{-1}(T,\mu_u,\mu_d,\sigma,\vec{\pi},A)=\nonumber\\
	&&i\frac{1}{\pi^2}\int_0^{\infty}\textrm{d}p\ p^2\left(\frac{\sin\lambda_1}{\cos\lambda_1+\cosh(\ve(p)/T)}+\frac{\sin\lambda_2}{\cos\lambda_2+\cosh(\ve(p)/T)}+\frac{\sin\lambda_3}{\cos\lambda_3+\cosh(\ve(p)/T)}\right)
\ee
and
\be\label{eq:c2udmt}
	&&\frac{\partial^2}{\partial\mu_u\partial\mu_d}\ln\det  S^{-1}(T,\mu_u,\mu_d,\sigma,\vec{\pi},A)=\nonumber\\
	&&\Bigg[\frac{1}{\pi^2}\int_0^{\infty}\textrm{d}p\ p^2\Bigg(\sum_{i=1}^3\Bigg(\frac{1+\cos\lambda_i(\cosh(\ve(p)/T)}{T\ve^2(p)(\cos\lambda_i+\cosh(\ve(p)/T))^2}+\frac{\cos\lambda_i}{\ve^3(p)(\cos\lambda_i+\cosh(\ve(p)/T))}\nonumber\\
	&&+\frac{\cosh(\ve(p)/T)-\sinh(\ve(p)/T)}{\ve^3(p)(\cos\lambda_i+\cosh(\ve(p)/T))}\Bigg)\Bigg)-\frac{1}{\pi^2}\int_0^{\Lambda}\textrm{d}p\frac{3p^2}{\ve(p)^3}\Bigg]\pi^+\pi^-
\ee
where $\varepsilon(p)=\sqrt{p^2+(m_0-\sigma)^2+\vec{\pi}^2}$, while the $\lambda_i$ are the eigenvalues of the Polyakov loop matrix:
\be
	\lambda_1=\dfrac{A_3}{T}+\dfrac{A_8}{\sqrt{3}T},\  &\lambda_2=-\dfrac{A_3}{T}+\dfrac{A_8}{\sqrt{3}T},\  &\lambda_3=-\dfrac{2A_8}{\sqrt{3}T}.
\ee

From Eq.~(\ref{eq:c2udfd}) we observe that this expression is odd with respect to $A_8$ and even in all other fields. As a consequence the expectation value of such a term is zero when the functional integration on the domain of $A_8$ is performed. This implies that the third term in Eq.~(\ref{eq:c2ud}) vanishes.

Another crucial observation is that taking the mean field limit for the pion field, $\vec{\pi}=\langle\vec{\pi}\rangle=0$, Eq.~(\ref{eq:c2udmt}) vanishes. Consequently, $\chi_{ud}$ vanishes altogether in the mean field approximation, independent of the temperature. Computing the expectation values of Eqs.~(\ref{eq:c2udfd}) and (\ref{eq:c2udmt}) using the MC-PNJL approach we include corrections induced by fluctuations of the pionic and Polyakov loop fields. Moreover, the main contribution to the Eq.~(\ref{eq:c2udmt}) is given by the pionic fluctuations, whereas the Eq.~(\ref{eq:c2udfd}) is non-zero mostly due to fluctuations of $A_8$.

\begin{figure}[ht!]
       \centering
        \subfigure{\includegraphics[width=.45\textwidth]{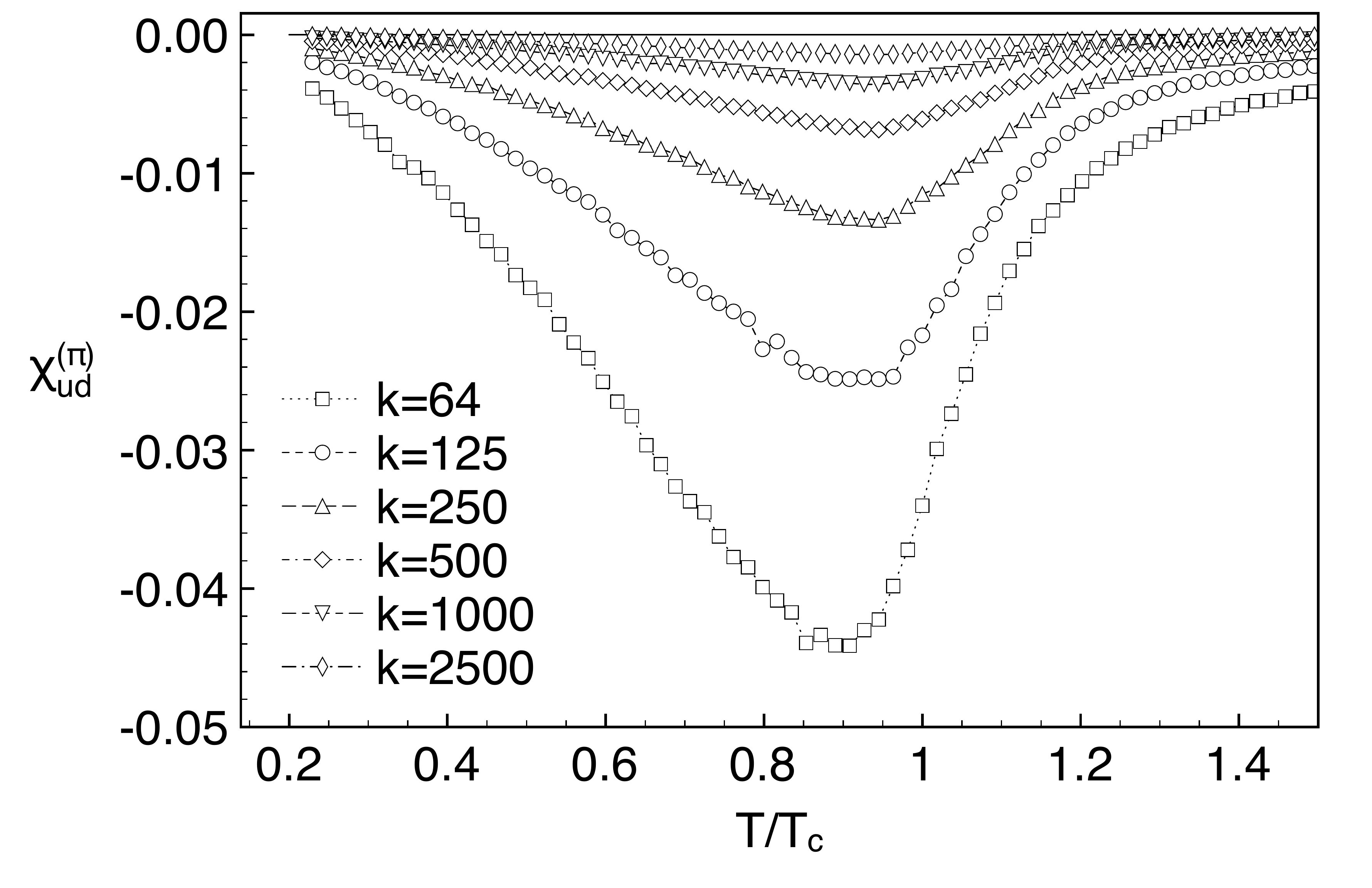}}\hspace{5mm}%
       \subfigure{\includegraphics[width=.45\textwidth]{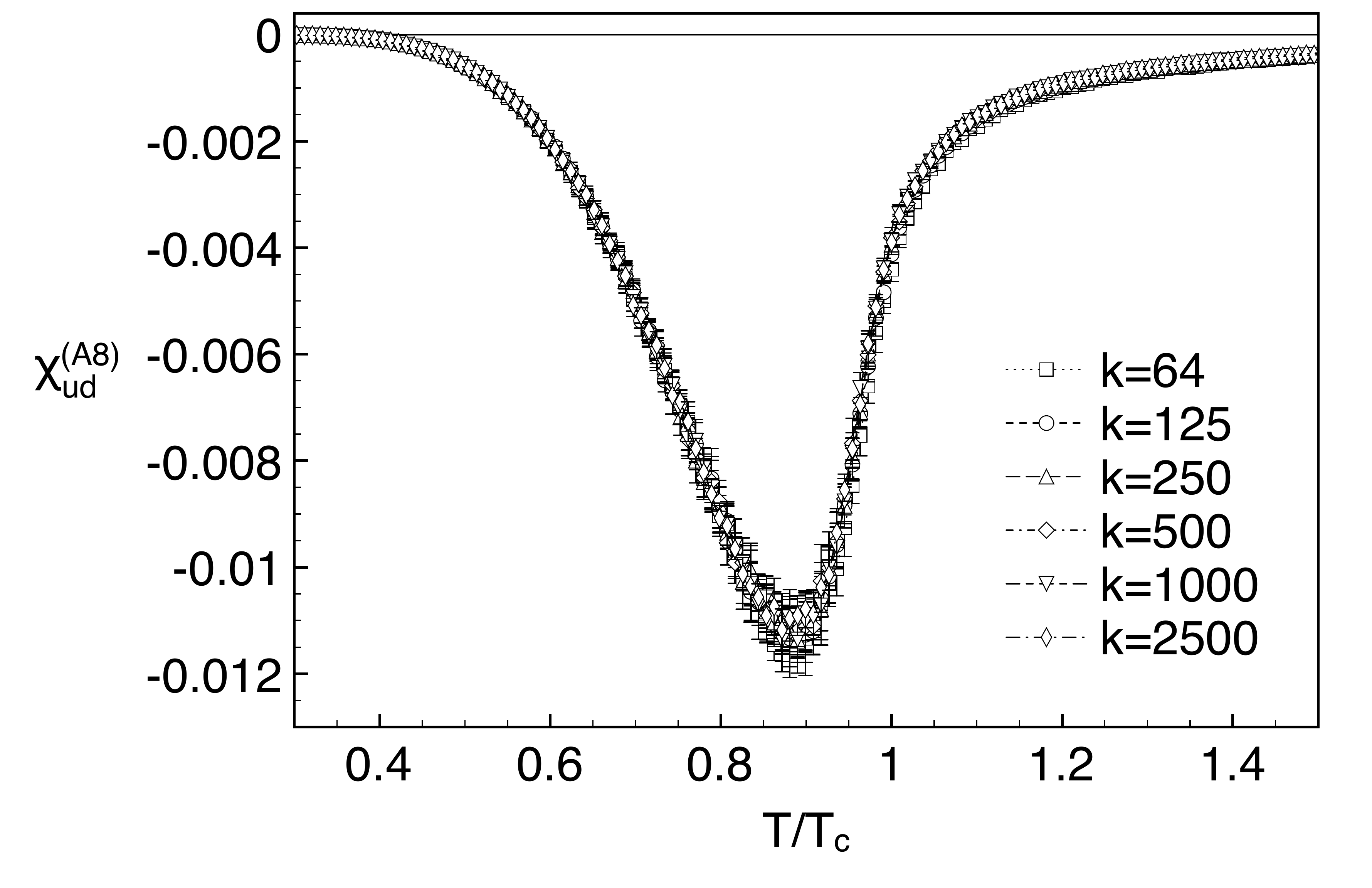}}\hspace{5mm}%
      \caption{Different contributions to the off-diagonal susceptibility $\chi_{11}=\chi_{ud}$ for different volume ratios $k$ computed in the Monte-Carlo approach. Left panel: Contribution from pionic fluctuations, for which the volume dependence is large. Right panel: Contribution from fluctuations of the $A_8$ field, which show a negligible volume dependence.}\label{fig:c2udD1D2}
\end{figure}

The pionic and $A_8$ contributions to $\chi_{ud}$ resulting from the MC-PNJL computation are shown in Fig.~\ref{fig:c2udD1D2}. Two characteristic features are immediately apparent. First, the term involving pionic zero-modes is strongly volume dependent and vanishes in the limit of infinite volume. Secondly, the term associated with fluctuations of the $A_8$ gauge field is independent of the box size and survives in fact as the volume becomes infinitely large.

\subsection{Chiral effective Lagrangian}

In order to better understand the role of the pionic fluctuations in the evaluation of $\chi_{ud}$, let us briefly digress and study this issue in the context of chiral perturbation theory (ChPT). 

For low temperatures and small values of the chemical potential, the physics is dominated by the effects of light pions and we can describe the system in terms of an effective chiral Lagrangian. While not influenced by the baryon number chemical potential, the pions do couple to the isovector chemical potential and its effect can be included in this effective Lagrangian. The chemical potential enters into the QCD Lagrangian like the zeroth component of a gauge field \cite{Kogut:1999iv,Kogut:2000ek}. When one promotes the global chiral flavor symmetry $SU(2)_L \times SU(2)_R$ of the QCD Lagrangian to a local symmetry, gauge invariance determines completely how the chemical potential must be implemented in the effective chiral Lagrangian \cite{Son:2000xc,Son:2000by,Splittorff:2002xn}. This Lagrangian has the form, expressed in terms of the chiral field $\Sigma$,
\be
{\mathcal L} &=& \frac{f_\pi^2}{4} \mathrm{Tr}\left[\nabla_\nu \Sigma \nabla_\nu \Sigma^\dagger\right] -\frac{m\bar{\Sigma}}{4} \mathrm{Tr}\left[\Sigma + \Sigma^\dagger\right],
\label{eq:cheff}
\ee
where $\bar{\Sigma} = |\langle \bar{\psi}\psi\rangle|$ is the magnitude of the chiral condensate and the covariant derivatives are defined as
\be
&\nabla_0\Sigma = \partial_0 \Sigma + \mu_I\left[\tau_3 \Sigma - \Sigma \tau_3\right], \quad \quad &\nabla_i \Sigma = \partial_i \Sigma, \quad\quad i=1,2,3\nonumber\\
&\nabla_0\Sigma^\dagger = \partial_0 \Sigma^\dagger + \mu_I\left[\tau_3 \Sigma^\dagger - \Sigma^\dagger \tau_3\right], \quad \quad
&\nabla_i\Sigma^\dagger = \partial_i \Sigma^\dagger, \quad i=1,2,3.
\ee
$\tau_3= \mathrm{diag}(1, -1)$ is the diagonal isospin generator.
Inserting these expressions into the chiral effective Lagrangian (\ref{eq:cheff}) allows to identify the terms depending on the isovector chemical potential:
\be
{\mathcal L} &=& \frac{f_\pi^2}{4} \mathrm{Tr} \left[\partial_\nu \Sigma \partial_\nu \Sigma^\dagger \right] + \frac{f_\pi^2}{2} \mu_I\mathrm{Tr} \left[(\partial_0 \Sigma)\Sigma^\dagger \tau_3 + \Sigma^\dagger (\partial_0 \Sigma)\tau_3\right] \nonumber\\
&&+ \frac{f_\pi^2}{2} \mu_I^2 \mathrm{Tr} \left(\tau_3 \Sigma \tau_3 \Sigma^\dagger - \openone_2\right) - \frac{m\bar{\Sigma}}{4} \mathrm{Tr}\left[\Sigma + \Sigma^\dagger\right]
\ee

For $|\mu_I | \ge m_\pi/2$, the formation of a pion condensate is possible and the ground state changes from the one at $\mu_I=0$, which is determined solely by the chiral condensate, to a combination of chiral and pion condensates \cite{Son:2000xc,Son:2000by,Klein:2003fy}. Since we are interested in the determination of the susceptibility $\chi_{ud}$ defined in the limit $\mu_I=0$, we can expand around the unchanged ground state and the chiral field $\Sigma \in SU(2)$ can be parameterized as
\be
\Sigma = \exp\left(i \frac{\pi^a}{f_\pi} \tau^a \right),
\ee 
where the $\tau^a$, $a=1, 2, 3$ are the generators of $SU(2)$ with normalization $\mathrm{Tr}(\tau^a \tau^b)= 2 \delta^{ab}$. Expanding the Lagrangian to second order in the pion fields $\pi^a$, one finds
\be
\mathcal{L} &=& \frac{1}{2} (\partial_\nu \pi^a)(\partial_\nu \pi^a) + i 2 \mu_I f_\pi (\partial_0 \pi^3) + i 2 \mu_I \left[(\partial_0 \pi^1)\pi^2 - (\partial_0 \pi^2) \pi^1\right] \nonumber\\
&&+\frac{1}{2}m_\pi^2 \pi^a \pi^a - 2 \mu_I^2 (\pi^1\pi^1 + \pi^2 \pi^2)
\ee
where we have identified $m\bar{\Sigma} = f_\pi^2 m_\pi^2$. In order to make contact with the results from the Monte-Carlo evaluation of the PNJL model with fluctuations of the mean-field only, it is sufficient to take the static part of the Lagrangian into account. The partition function for this static part is 
\be
\mathcal{Z}_{\mathrm{static}} &=& \int \prod_{a=1}^{3} d \pi^a \exp\left\{-\frac{V}{T}\left[\frac{1}{2} m_\pi^2 \pi^a \pi^a - 2 \mu_I^2 (\pi^1\pi^1 + \pi^2 \pi^2) \right]\right\}\nonumber\\
	&=&  \int \prod_{a=1}^{3} d \pi^a \exp\left\{-\frac{V}{T}\left[\frac{1}{2} m_\pi^2 \pi^a \pi^a -  4\mu_I^2 (\pi^+\pi^-) \right]\right\}.
\label{Zstat}
\ee 

\begin{figure}[ht!]
       \centering
      \includegraphics[width=.6\textwidth]{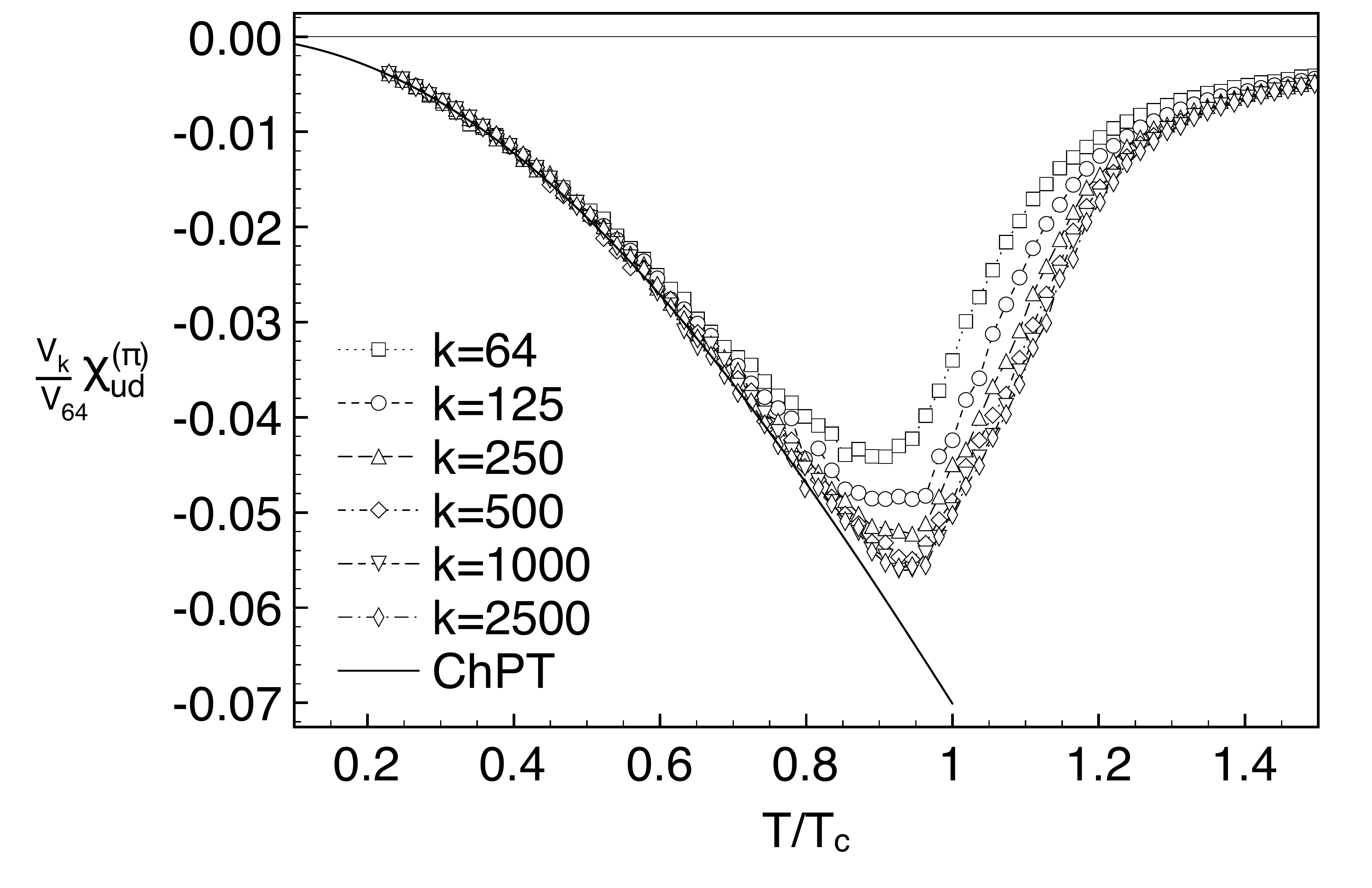}
      \caption{Scaled pionic contribution to the off-diagonal susceptibility compared with the ChPT prediction. All Monte-Carlo PNJL results are multiplied by the volume $V_k/V_{64}$ and therefore scale with the $k=64$ curve, using Eq.~(\ref{eq:chptrel}).}\label{fig:c2chpt}
\end{figure}

The second-order off-diagonal expansion coefficient $\chi_{ud}$ is given by
\be
\chi_{ud}^{(\pi)} &=& \frac{1}{VT} \left.\frac{\partial^2}{\partial\mu_u \partial\mu_d} \ln \mathcal{Z}_{\mathrm{static}}\right|_{\mu_u=\mu_d=0}
\ee
and receives contributions from the two-point correlations $\langle \pi^+\pi^-\rangle$. Evaluating the integral, we obtain the static result for $\chi_{ud}$ from the chiral effective Lagrangian:
\be\label{eq:chptrel}
\chi_{ud}^{(\pi)} &=& -\frac{2}{V T m_\pi^2} = -\frac{2\ T^2}{k}\frac{1}{m_\pi^2}, 
\ee
setting again $V=k/T^{3}$. 
This prediction can be compared directly with our Monte-Carlo PNJL results, provided we take the temperature dependence of the pion mass into account, using the relation given in \cite{Gasser:1986vb}:
\be
	&&m_{\pi}(T)=m_{\pi}\left(1+\frac{g_1(m_{\pi}^2,T,L)}{4f_{\pi}^2}+\mathcal{O}(p^4)\right),\\
	&&g_1(m_{\pi}^2,T,L)=\frac{1}{(4\pi)^2}\int_0^{\infty}\textrm{d}\lambda^{r-3}\sum_{n\neq0}\exp(-m_{\pi}^2\lambda-n^2/(4\lambda)),\nonumber\\
	&&n=(n_1L,n_2L,n_3L).
\ee
Fig.~\ref{fig:c2chpt} demonstrates that the picture so obtained from the chiral effective Lagrangian is completely consistent with our Monte-Carlo calculations in the PNJL model, as far as the pionic contributions to $\chi_{ud}$ are concerned.
From Fig.~\ref{fig:c2chpt} it also follows that the chiral perturbation theory prediction for this coefficient is reliable until around $T/T_c\simeq0.7$. 

\subsection{Comparison with lattice data}
\begin{figure}[ht!]
       \centering
      \includegraphics[width=.6\textwidth]{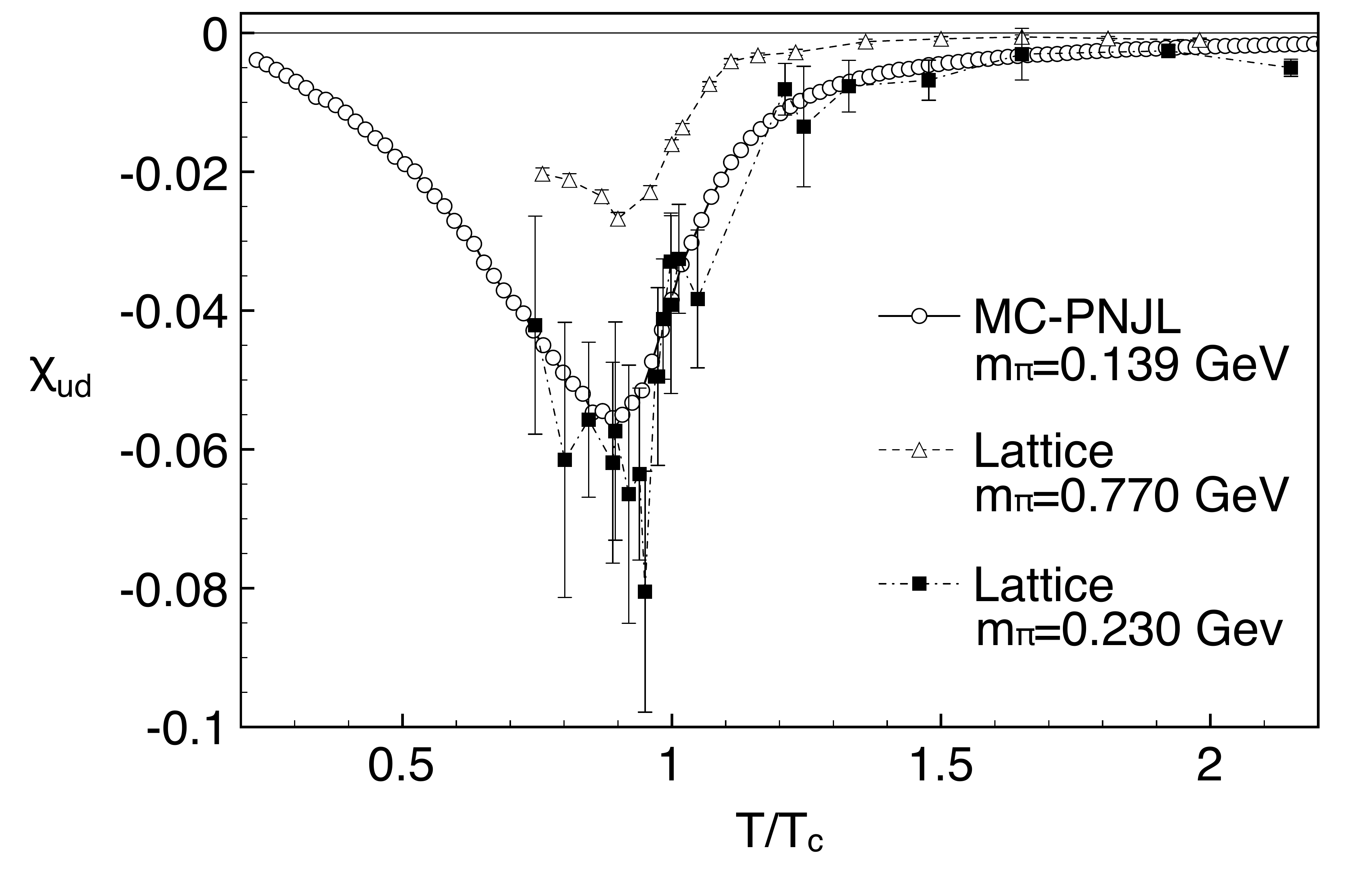}
      \caption{Temperature dependence of the flavor off-diagonal susceptibility $\chi_{ud}$ in the Monte-Carlo approach to the PNJL model, using $k=64$ ($LT=4$). Lattice data \cite{Allton:2005gk,Gavai:2008zr} with the same volume aspect ratio $LT$ and different pion masses are also shown for orientation.}\label{fig:c2udtot}
\end{figure}

Lattice QCD studies of $\chi_{ud}$ have been carried out for example in Refs.~\cite{Allton:2005gk,Gavai:2008zr}, both with $k=64$ but with different quark masses, corresponding to pion masses $m_{\pi}=230$~MeV and $770$~MeV. These lattice results are compared to our Monte-Carlo PNJL computations (using the physical pion mass) in Fig.~\ref{fig:c2udtot}. The shape of the $\chi_{ud}$ signal is quite well reproduced within the large error band of the lattice data. The difference between lattice results computed with different pion masses is now quite plausible. Given that the pionic fluctuations dominate over those from the $A_8$ component of the Polyakov loop field, this behavior is just what one expects from Eq.~(\ref{eq:chptrel}). At the same time one would expect that lattice simulations performed ideally with physical quark masses would actually yield even larger magnitudes of $\chi_{ud}$ than those with $m_{\pi}=230$~MeV. The Monte-Carlo results notably include only the pionic zero modes. Finite-momentum fluctuations would tend to further increase the pionic effects in $\chi_{ud}$.

In the infinite-volume limit, only the gauge field signal (plus possible finite-momentum pionic modes) survives in $\chi_{ud}$, as pointed out earlier. It would be interesting to see whether this predicted behavior is realized in lattice QCD when moving towards larger values of the ratio $N_s/N_t$.

\section{Conclusions}\label{sec:concl}
Spontaneous chiral symmetry breaking and confinement are well-known phenomena emerging in the study of QCD thermodynamics. Both these features are incorporated in the Polyakov-loop extended Nambu--Jona-Lasinio (PNJL) model. Predictions for the deconfinement and the chiral restoration transition temperatures, obtained from mean-field calculations in this framework, compare quite well with the available lattice data. Nevertheless, the mean-field approximation is not sufficient if one wants to correctly account for the temperature dependence of other observables such as the Taylor expansion coefficients of the pressure in powers of $u$-and $d$-quark chemical potentials.

In this work we have applied standard Monte-Carlo techniques to a PNJL model in order to go beyond the saddle-point approximation. This becomes important when the system is considered in a finite volume. The strength of the fluctuations introduced in this way depends on the size of the volume. At fixed temperature $T$, the spatial volume size $L=V^{1/3}$ is determined by the factor $k^{1/3}=LT$ of the Euclidean volume.
We have checked the method by studying the chiral susceptibility for different volume sizes $V$ and shown that the saddle-point result is approached in the limit $V\rightarrow\infty$. Studying the thermodynamics, we found that the introduction of mean-field fluctuations in a finite volume does not change the traced Polyakov loop, both in the pure gauge sector and for $N_f=2$ quark flavors. The chiral condensate with $N_f=2$ experiences modest changes through such effects. 

On the other hand, the inclusion of such beyond mean field fluctuations in a finite volume does affect the susceptibilities significantly. We find that their impact is crucial for the evaluation of higher-order Taylor expansion coefficients of the pressure. In particular, the second-order flavor non-diagonal expansion coefficient $\chi_{ud}$ becomes non-zero. Our result from a Monte-Carlo computation agrees well with lattice data using the same $k$ for the Euclidean volume. In contrast, this coefficient vanishes in the saddle-point approximation. Good agreement with lattice results is also found for the second diagonal moment of the pressure. These results show that finite-volume fluctuations with pion quantum numbers can account for the non-vanishing expectation value of the susceptibility $\chi_{ud}$. This zero-mode contribution vanishes in the infinite-volume limit. Concerning the role of the pionic zero-mode fluctuations, it is also demonstrated that the Monte-Carlo results for $\chi_{ud}$ are fully consistent with those from chiral perturbation theory for temperatures below $T_c$. 

The contribution from the difference in the expectation values of the Polyakov loop and its conjugate, which also contributes to $\chi_{ud}$ through the fluctuations of the $A_8$ component of the gauge field, remains finite even in the limit $V\to \infty$. In lattice simulations, both types of mean-field fluctuations (pionic and gauge fields) contribute to the observed susceptibilities, although the pionic effects are expected to be smaller for the larger pion masses used in lattice QCD. While these pionic fluctuations will be suppressed for large volumes (i.e. a large $LT$ at fixed temperature), gauge field effects can still account for a non-zero $\chi_{ud}$ even in the infinite-volume limit. The present analysis predicts a decrease of $\chi_{ud}$ for larger volume sizes in lattice simulations. 

In conclusion, we have shown how finite-volume fluctuations in the PNJL model at non-zero chemical potential can be studied using a Monte-Carlo evaluation of the partition function, and that such fluctuations can affect susceptibilities significantly. Going beyond the mean-field approximation for the investigation of critical behavior requires additionally to take fluctuations on all momentum scales into account, and such a calculation would also be amenable to a Monte-Carlo evaluation. 

\begin{acknowledgments}
Stimulating discussions with Kenji Fukushima are gratefully acknowledged.
\end{acknowledgments}

\bibliography{PNJL}
\end{document}